# Electric-induced reversal of morphogenesis in *Hydra*

Erez Braun[1,*] and Hillel Ori[2]

[1]Department of Physics, [2]Faculty of Medicine,

Technion-Israel Institute of Technology, Haifa 32000, Israel.

[*]To whom correspondence may be addressed. Email: erez@physics.technion.ac.il

# ABSTRACT

**Morphogenesis involves the dynamic interplay of biochemical, mechanical and electrical processes. Here we ask: to what extent can the course of morphogenesis be modulated and controlled by an external electric field? We show that above a critical amplitude, an external electric field can halt morphogenesis in *Hydra* regeneration. Moreover, above this critical amplitude, the electric field can even lead to reversal dynamics: a fully developed *Hydra* folds back into its incipient spheroid morphology. The potential to renew morphogenesis is re-exposed when the field is reduced back to amplitudes below criticality. These dynamics are accompanied by modulations of the *Wnt3* activity, a central component of the head organizer in *Hydra*. Reversal of morphogenesis is shown to be triggered by enhanced epithelial electrical excitations, accompanied by intensified calcium activity, indicating that electrical processes play an instructive role to a level that can direct developmental trajectories. Reversal of morphogenesis by external fields, calls for extending its framework beyond programmatic, forward-driven, hierarchical processes.**

**Introduction**

Morphogenesis results from the dynamic interplay of three types of processes: biochemical [1-3], mechanical [2, 4] and electrical [5, 6], spanning all scales from the molecular to the entire organism [7, 8]. The robustness of the morphogenetic process is typically attributed to the presence of a well-defined hierarchy of forward-driven processes which lead to the emergence of a viable organism in a program-like manner [9]. The nature of the underlying mechanisms, such as threshold-crossing cellular processes and the development of symmetry-breaking fields, is considered as the source of stability of the emerging body plan. Yet, at the level of a cell, it has become clear that cellular differentiation, once considered to be highly stable and robust, can in fact be modulated and directly reprogrammed on demand [10]. Is it possible to 'reprogram' an entire multicellular organism? Is it possible to modulate the course of morphogenesis in a whole animal and alter its developmental trajectory in a controlled manner? Currently these questions remain open and there are no known examples of controlled reprogramming at the level of a whole organism.

Direct control of the course of morphogenesis in a non-destructive way is an outstanding challenge. The implementation of such control sets stringent demands on the system; the external control must be strong enough to induce a severe perturbation, yet the system has to maintain tight coordination of the underlying biochemical, mechanical and electrical processes and retain its viability. The ability to control morphogenesis will provide novel insights into the process of morphogenesis as well as advance practical applications in regenerative medicine and tissue engineering. We sought to develop such a control and utilize external electric fields, which can be exquisitely controlled and manipulated experimentally, to modulate the course of development in whole-body *Hydra* regeneration. Here we demonstrate that whole-body morphogenesis can indeed be controlled and even reversed, by the utilization of an external AC electric field.

Historically, studies of morphogenesis concentrated mainly on biochemical patterning by processes of reaction-diffusion and cell differentiation due to positional information [1]. The concept that developmental trajectories are determined in a program-like manner was established under the view that biochemical patterning plays a leading role in the dynamics. It is becoming evident that mechanical and electrical processes can also have a central role in morphogenesis. The idea that mechanical processes are important goes back to the seminal work of D'Arcy Thompson [11]. More recently, with the advance of technological capabilities, experimental efforts have started to shed light on the coupling between mechanics and bio-signaling during morphogenesis, showing multidirectional cross talks between these two type of processes [12].

Electrical processes also play important roles in morphogenesis; it has been demonstrated that endogenous voltage gradients drive wound healing in animals (including humans) and plants [13]. The electrical processes in cells (even non-excitable) seem to play a regulatory role and are integrated during embryogenesis with the biochemical signals [14]. The connection between bioelectric processes and other processes leading to cell proliferation, motility and apoptosis in morphogenesis during development and cancer is beginning to be elucidated [13, 15].

Perturbing the non-neural bioelectric signals in embryogenesis and regeneration can be utilized by drugs affecting the underlying molecular machinery (e.g., gap junctions and ion channels). Using these methods, it has been shown that electrical processes regulate gene expression and participate in organ morphogenesis as well as in patterning the body plan at different levels in various developmental and regeneration model systems. Modulations of these electrical processes lead to modifications in patterning and ectopic structure formation. Bioelectricity participates and even in some cases directs patterning in embryogenesis and regeneration. For example, manipulation of specific ion channels in frogs induces eye formation in many different tissues, such as the gut and the tail [14]. Altering the electrical properties of the tissue in *Planaria* regeneration, causes dramatic modifications in morphogenesis by affecting polarity and leading to the emergence of animals with multiple heads. Temporary reduction of the bioelectric connectivity in *Planaria* tissue during regeneration also induces head formation of different morphologies (see recent reviews summarizing these phenomena [5, 6, 16]).

Attempts to elucidate the role of electrical processes in development also involved the utilization of external electric fields. These attempts have a rich history going back to Roux, who applied external electric fields to developing eggs already in the 19${}^{th}$ century [17]. More recently, applications of external electric fields showed that different cell types respond to the external field in a positional sensitive manner, leading to directional cell migration and cell proliferation affecting morphogenesis [13]. In animals, it has been demonstrated that the application of electric fields leads to regeneration of limb form, including muscle, nerve and cartilage in amputated forelimbs of animals which do not normally regenerate them, such as *Xenopus* and rats [13]. External electric fields were also utilized to demonstrate that they can affect the polarity of the body axis via directing charged morphogens [18].

Regeneration of a whole animal from a tissue segment provides a powerful model to study morphogenesis due to its flexibility, allowing to apply a wide range of experimental manipulations. Regeneration processes are closely related and utilize similar molecular components to embryogenesis, so studies of these different developmental processes complement each other [12, 19].

Among models of whole-body regeneration, *Hydra*, a freshwater animal, stands out in its remarkable regeneration capabilities [20]. *Hydra* played a crucial role in the history of experimental biology thanks to the studies of Abraham Trembley in the 18$^{th}$ century, who discovered regeneration and demonstrated that bisected *Hydra* can regenerate a head or a foot according to their original polarity [21, 22]. It was later shown that even small excised tissue segments or condensed aggregates formed from a mixture of dissociated cells, regenerate into complete animals within a couple of days [20, 23]. A tissue segment first goes through an essential stage in which it forms a hollow spheroid made of a bilayer of epithelial cells which eventually regenerates into the body of a mature *Hydra* [24-26].

*Hydra* also possesses outstanding electrical properties, making it a natural model system to study the effect of external electric fields on morphogenesis. Practically, all cells in *Hydra* are electrically excitable and the epithelial tissues are capable of generating and propagating electrical action potentials [27-30]. This means, that the outer epithelium layer does not present a passive barrier to external electrical perturbations; even relatively weak external electric fields can trigger strong internal effects due to the non-linear amplification by the generation of excitable electrical spikes in the tissue, and can thus trigger strong effects.

Here we show that an external electric field can be tuned to halt the process of morphogenesis in a non-destructive way in *Hydra* regeneration. Moreover, the external electric field can be further adjusted to drive morphogenesis backward and forward, around a critical point in a controlled manner. To the best of our knowledge, such a remarkable level of control on the course of morphogenesis by an external field has never been demonstrated before. In particular, a backward-forward cycle of morphogenesis leads to a newly emerged body plan in the re-developed folded tissue, which is not necessarily similar to the one before the reversal process. Thus, a controlled drive of morphogenesis allows in principle, multiple re-initiation of novel developmental trajectories for the same tissue.

We show that the above phenomena are mediated by enhanced electrical excitations accompanied by intensified calcium dynamics. Thus, the external electrical field stimulates a response which is amplified by the excitable machinery of the *Hydra* tissue showing that electrical processes play instructive role in morphogenesis. We further show that the expression level of *Wnt3*, a central signaling component in *Hydra,* is modulated to first decay and then turned on, respectively with the cycle of backward and forward morphogenesis controlled by the external field. These observations demonstrate the tight coordination of the biochemical and mechanical processes with the electrical ones, ensuring the integrity of the tissue and its regeneration capability as morphogenesis is folded back.

**Results**

*Morphology dynamics*

We place small *Hydra* tissue spheroids between a pair of platinum mesh electrodes and study the regeneration dynamics under applied electric fields (Methods). In the absence of external electric fields, *Hydra* regenerate from incipient tissue spheroids into mature animals within 15–55 hrs [25]. We find that the regeneration trajectory is significantly affected by the application of an external electric field (Fig. 1). The application of an AC voltage above a critical amplitude of ~20–30 Volts at 1 kHz frequency, halts regeneration and the tissue does not develop (Figs. 1a,b; Fig. S1). The suspended tissue maintains its regeneration potential as proved by the resumption of morphogenesis upon the reduction of the external voltage below the critical value (Fig. 1b;

Fig. S1). When the voltage is increased above the critical value after the regeneration process concluded, a fully developed *Hydra* gradually shrinks its tentacles and eventually folds back its mature body-plan into the incipient morphology of a spheroid (Fig. 1c; Fig. S2).

The reversal of morphogenesis is gradual and can be controlled by the external voltage however, different tissues exhibit different critical voltages. This imposes a practical challenge to precisely controlling the morphogenesis trajectory, due to the individual sensitivity of each tissue to the applied voltage. When the voltage increase is gradual and carefully controlled near the critical value, the tissue maintains its integrity as well as its regeneration capacity for repeated cycles of backward-forward morphogenesis (Supplementary Movie 1). This reversal of morphogenesis is reproducible and robust as summarized in the cumulative distributions of Fig. 1d. Below the critical voltage (blue curve, left) regeneration (defined by the emergence of clearly observable tentacles) emerges for all samples in less than ~55 hrs (147 tissue samples from 15 different experiments). At voltages above criticality, the probability to observe tentacles in the back-folding samples diminishes within a wide range of time scales, and is practically zero above ~90 hrs. All 77 tissue samples showing reversal of morphogenesis (from 15 different experiments) in Fig. 1d (red curve, right) first regenerate into mature animals and then fold back upon the increase of the external voltage.

Renewed morphogenesis upon the reduction of the applied voltage, does not necessarily lead to the same morphology as in the previous cycle, e.g., manifested by a different body form, or number of tentacles and their shape (Fig. S2). This is indicative that the folding back of the morphology is not a transient stress response but rather involves significant modulations of the tissue organization. When the cycle of controlled backward-forward morphogenesis is repeated a few times, the subsequent cycle of backward folding requires higher voltage to reverse morphogenesis than the previous one (see examples in Fig. 4a; Fig. S2). This increase in the critical voltage points to an interesting adaptation of the tissue.

*Bio-signaling dynamics*

Does reversal morphogenesis have any signature in the underlying biological processes beyond the observed change in morphology? We utilize a transgenic *Hydra* expressing a GFP probe under the control of the *Wnt3* promoter [31, 32]. *Wnt3* is a component of the canonical *Wnt* pathway, which has been shown to be key in patterning the body plan in many organisms including *Hydra* [31, 33]. *Wnt3* has been shown to be a critical component of the *Hydra* head organizer, which is a well-defined group of cells. *Wnt3* is continuously maintained active by an autoregulatory transcriptional system [31] and plays an important role in preserving the integrity of the body-plan in face of continuous replacement of tissues in a developmental process that never ceases [34]. The *Wnt3* signal is thus a relevant marker for the underlying biological state of the tissue in morphogenesis.

We find that, the *Wnt3*-activity fluorescence signal that emerges at the tip of the head (hypostome) in mature *Hydra* [31, 33], gradually decays after the external voltage is increased above criticality, marking the decay of the head organizer activity as the tissue folds back into an incipient spheroid (Fig. 2; Fig. S3). The *Wnt3* activity is estimated from time-lapse microscopy, measuring the fluorescence density in a small area around the GFP signal at the head, normalized by the average fluorescence density of the tissue outside this region. The *Wnt3* signal re-emerges upon a second round of regeneration, following a switch of the external voltage back to zero (Fig. 2 and Fig. S3; Supplementary Movie 2).

Overall, we analyzed in detail 7 tissue samples (from 5 different experiments) of fully regenerated *Hydra*, reversing their morphology together with the decay of the *Wnt3*-activity fluorescence signal (Fig. 2 and Fig. S3). The inset to Fig. 2 shows the mean *Wnt3*-activity decay as a function of time at external voltages above the critical values, measured from the GFP fluorescence density levels averaged over these 7 tissue samples. Since the critical voltage and correspondingly the experimental protocol of the applied voltage is different for each sample, the fluorescence trace was first shifted to zero time at the peak of the fluorescence signal and then interpolated to give a uniform sampling coverage for all traces before averaging. As shown by the shaded area around the curve, the standard error of this averaging is small. Thus, the significant *Wnt3* decay observed is a direct response of the bio-signaling system to the external voltage. Note that this decay of the fluorescence signal indeed marks the decay of the *Wnt3* activity, as the latter is transcriptionally controlled by its autocatalytic activity [31].

*Calcium dynamics and Electrical processes*

To gain insight into the mechanism of reversible morphogenesis, we turn to study the effect of the external field on calcium dynamics. Calcium is a natural candidate as it is a universal effector across biological systems and an important mediator between mechanical, electrical and biochemical processes [12]. Towards this end, we constructed a transgenic *Hydra* expressing a fast $Ca^{2+}$ fluorescence probe in its epithelial cells [30, 35] (Methods). Tissue fragments excised from these transgenic *Hydra* are placed in the experimental setup after folding into spheroids and time-lapse movies at 1 min resolution are recorded. The 1 min time resolution enables a measurement of the entire regeneration process, possibly for more than one cycle of backward-forward morphogenesis, over days without significant damage to the tissue. While higher temporal resolution measurements could reveal information on the $Ca^{2+}$ kinetics [30], the minute resolution enables to measure the *average* modulations of the $Ca^{2+}$ activity under the external field for extended periods.

Altogether we studied 9 tissue samples (from 3 different experiments) and collected more than 10,000 measurement points in order to uncover Ca dynamics in regeneration under modulated electric fields. Individual examples and cohort statistics are shown in Figure 3. In the absence of external electric fields, the tissue initially exhibits local excitations of the $Ca^{2+}$ signals at different parts, which eventually become coordinated into whole-tissue coherent spikes (examples in Figs. 3a,b). Elevated external voltage (>15 V, 1 kHz), leads to a significant increase in the $Ca^{2+}$ activity by an increased density of the $Ca^{2+}$ spikes train riding on an enhanced baseline. Even more enhanced activity is observed above a critical voltage of ~20 V (Fig. 3a, arrow at the lower trace), with prolonged periods showing almost continuous enhanced activity. All fluorescence traces, measure the fluorescence density (fluorescence per unit area) averaged over the entire tissue relative to the mean fluorescence density in the background (which is much lower than that of the tissue; see Supplementary Movie 3).

Applying an external voltage immediately following the folding of the excised tissue to a spheroid, Fig. 3c (upper trace; Fig. S4a) depicts the fluorescence density of the initial spheroid tissue under 15 V (1 kHz). At this voltage, which is somewhat below the critical value, the tissue immediately exhibits a significant level of $Ca^{2+}$ activity, which apparently is not enough to halt morphogenesis (see attached images, Fig. 3d; Fig. S4a). Increasing the voltage above criticality, further increases the level of $Ca^{2+}$ activity (Fig. 3c, middle trace; Fig. S4b) leading in turn to the folding of the already patterned *Hydra* back into the incipient spheroid (upper image in Fig. 3e; Fig. S4b). Finally, switching the voltage to zero reduces $Ca^{2+}$ activity (Fig. 3c, bottom trace; Fig. S4c) and leads to the recovery of morphogenesis and renewed regeneration of a mature *Hydra* (lower image in Fig. 3e; Fig. S4c), indicating that the tissue maintains its regeneration potential (Supplementary Movie 3). The normalized distributions of fluorescence densities in Fig. 3f (9 tissue samples from 3 different experiments, >10,000 measurement points for each curve) quantify a statistical measure of the $Ca^{2+}$ activity, showing a significant enhanced tail of the distribution at elevated external voltages (red) compared to the one at zero voltage (blue).

We next show that the elevated $Ca^{2+}$ activity leading to reversal of morphogenesis is a manifestation of enhanced electrical excitability of the *Hydra* tissue. This is demonstrated along three different experimental lines. First, we demonstrate the existence of a frequency cutoff for the AC electric field, above which the tissue becomes insensitive to the applied field; morphogenesis proceeds normally and no reversal is observed. We then show that increase in the external potassium concentration in the medium leads to enhanced calcium activity as well as reversal of morphogenesis. Finally, we also utilize direct electrical measurements, showing that calcium spikes are stimulated by bursts of action potentials.

Fig. 4a shows $Ca^{2+}$ traces at two different frequencies of the external field (1 kHz and 3 kHz). The initial tissue spheroid first regenerates into a mature *Hydra* at V=0, showing a typical low level of $Ca^{2+}$ activity (Fig. 4a, first trace; Fig. S5a). Increasing the external voltage above the critical value (26.5 V) at 1 kHz, leads to a significant elevation in the level of $Ca^{2+}$ activity and folding of the mature *Hydra* back into a spheroid morphology (Fig. 4a, second trace; Fig. S5b). Increasing the AC frequency to 3 kHz while maintaining the same voltage amplitude, results in reduced $Ca^{2+}$ activity and re-emergence of a regenerated mature *Hydra* (Fig. 4a, third trace; Fig. S5c). The statistics showing the frequency sensitivity is summarized in Fig. 4b, depicting the normalized distributions (2 tissue samples, >3000 points each) of fluorescence density at zero and high voltage at 1 kHz (20-26.5 V) and 3 kHz (26.5 V). The $Ca^{2+}$ activity at 3 kHz is comparable to that at zero voltage, while the activity at 1 kHz is significantly enhanced.

Repeating this frequency-switching cycle, Fig. 4a (fourth trace) shows the re-emergence of an elevated $Ca^{2+}$ activity and backward folding of morphogenesis into a spheroid upon switching back to 1 kHz. This second cycle of backward folding however, requires higher critical voltage to reverse morphogenesis than the first one (V>30 V). Switching the frequency again to 3 kHz (at the same voltage amplitude) reduces $Ca^{2+}$ activity and leads to the emergence of a re-regenerated mature *Hydra* (Fig. 4a, bottom trace).

These data demonstrate the existence of a frequency cutoff around 1 kHz, above which the elevated $Ca^{2+}$ activity is reduced and morphogenesis is restored from its suspended state. This upper frequency cutoff is not a sharp cutoff at precisely 1 kHz. Nevertheless, further experiments demonstrate that at frequencies higher than approximately 1 kHz, the regeneration process is insensitive to the external electric field (Supplementary Movie 4). The ~1 kHz frequency cutoff, above which the tissue becomes "transparent" to the applied external voltage, strongly suggests a corresponding characteristic timescale of the order of a few milliseconds, which is consistent with the measured capacitance (RC) time constant of the *Hydra* tissue [36].

We next follow the tissue dynamics under an elevated external potassium ($K^+$) concentration which serves as a standard method to stimulate excitable tissues, as demonstrated also in *Hydra* [37]. Fig. 4c shows that increasing potassium concentration in the medium from 0.1 mM (normal medium; upper trace) to 1 mM (lower trace), indeed leads to elevated $Ca^{2+}$ activity (see also the normalized distributions in Fig. 4d; 2 tissue samples, ~4000 points each curve). Importantly, this is accompanied by folding back of the regenerated *Hydra*. The tissue seems to lose its regeneration capability following the high potassium treatment. Nevertheless, it is clear that elevated potassium excites the system and leads to reversal morphogenesis similar to that observed under an external electric field (Fig. S6).

We finally directly measure the spontaneous time-dependent electrical potential by inserting a silver-chloride electrode into a tissue segment embedded in a low-melting 2% agarose gel that damps its motion, while simultaneously recording fluorescence images of the $Ca^{2+}$ signal under the microscope. These measurements are done in the absence of an external field stimulation and due to their invasive nature, do not allow following the regeneration process. Nevertheless, The example trace in Fig. 4e shows that every $Ca^{2+}$ spike observed in the tissue is stimulated by a pre-burst of action potentials [28, 30, 38] (overall we observed 167/174 measured $Ca^{2+}$ spikes showing clear pre-bursts of action potentials leading them; see Figs. S7a-b for more examples).

These three pieces of evidence: frequency cutoff, excitations due to elevated potassium and direct electrical recordings, point to enhanced electrical activity of the *Hydra* tissue as the source of elevated $Ca^{2+}$ activity and thus to the mechanism triggering reversal morphogenesis. Previous works indeed showed that external electric fields, within the same range used in our experiments, cause enhanced electrical activity in mature *Hydra*, resulting in enhanced discharge of nematocytes [37].

**Discussion**

This work studies the biophysics of morphogenesis under external electric fields. The main finding is that above a critical field amplitude, morphogenesis is halted and can even be reversed in a controlled manner. The reversal trajectory maintains the integrity of the tissue and its regeneration capability, leading in some of the cases to a different body plan in the next round of regeneration. The time for re-regeneration is similar to the original regeneration time. The required significant re-regeneration time, together with the emergence of a new body plan, show that the folding back of morphology is not a temporal stress response. Our experiments indeed demonstrate that this phenomenon is due to a genuine reversal trajectory of morphogenesis, involving both the folding back of morphology as well as the decay of the *Hydra* central signaling system at the head organizer. The electrical excitations stimulated by the external electric field lead to enhanced calcium excitations in the epithelium tissue. The reversal of morphogenesis can be triggered by the electrical excitations themselves or by the enhanced calcium activity in the tissue, or by the combined effect of these two types of processes. Since the *Hydra* epithelium is basically a muscle connected by whole-animal supracellular actin fibers [25], one possibility is that enhanced calcium excitations lead to strong mechanical perturbations which in turn trigger the reversal process. The mechanism by which enhanced calcium and electrical activities redirect and modulate the course of morphogenesis, remains an important open issue for future work.

The observed phenomena of halting and reversing of morphogenesis by electrical excitations, re-tuned by the external field, have several important implications. First, morphogenesis is usually considered a hierarchical forward-driven process, in which each stage switches-on the next one until the completion of a body plan [9]. It was shown before that under metabolic stress *Hydra* can lose its tentacles and change its body morphology in a reversible way, but this phenomenon was not further investigated [39]. The observation that an external physical manipulation in the form of an electric field, by triggering enhanced electrical excitations of the tissue, can halt morphogenesis and even reverse it in a controlled way while maintaining the regeneration potential of the tissue, paints a picture of morphogenesis in which electrical processes play an instructive role to a level that can direct developmental trajectories not obeying this hierarchy [5].

Second, it shows that the electrical processes are tightly integrated with the underlying mechanical and biochemical ones [12]. Folding back of the body-plan of a fully developed animal, simultaneously with the decay of a central signaling system at the head organizer, while maintaining the regeneration capability of the tissue is highly non-trivial. It requires tight integration and coordination of the tissue electrical processes, stimulating the reversal dynamics, with mechanical (structural changes resulting in tissue folding) and biochemical (bio-signaling) processes, in order to keep the integrity of the backward folded tissue. In particular, the decay of the head organizer during reversal of morphogenesis demonstrates that changes in the tissue encompass more than mere morphology, and also involve significant rearrangements in the underlying biological processes. How deep these biological transformations are, e.g. in reprogramming differentiated cells into stem cells, remains an exciting question for future studies. The ability to stimulate reversal of morphogenesis by an external electric field opens up new ways to further study the symbiotic dynamics of these different processes in morphogenesis. In particular, the ability to halt morphogenesis at different time points and reverse its dynamic trajectory enables us to identify the types of information stored in the developing tissues, the reservoirs of ions and biomolecules and structural memories [25] instilled in the tissue and playing a role in the regeneration process. Furthermore, the ability to block regeneration in a controllable fashion can shed light on the origin of a tissue's regeneration potential and reveal the reasons some tissues can readily regenerate while others cannot do so.

Third, controlled reversal of morphogenesis opens the possibility for a new approach in the study of developmental systems: One would like to study the potential of a given system to realize different developmental trajectories, beyond the canalized one [12]. Currently, this can be done only by studying an ensemble of different systems. Controlled halting, reversing and re-initiating a developmental process of a given tissue at different time points on demand, open up a possibility for studying the developmental potential rather than an instantiation of it.

Overall, the methodology and phenomenology exposed in this work offer a unique access for studying the physics underlying one of the most fundamental processes in living systems, that is - morphogenesis.

**Figures:**

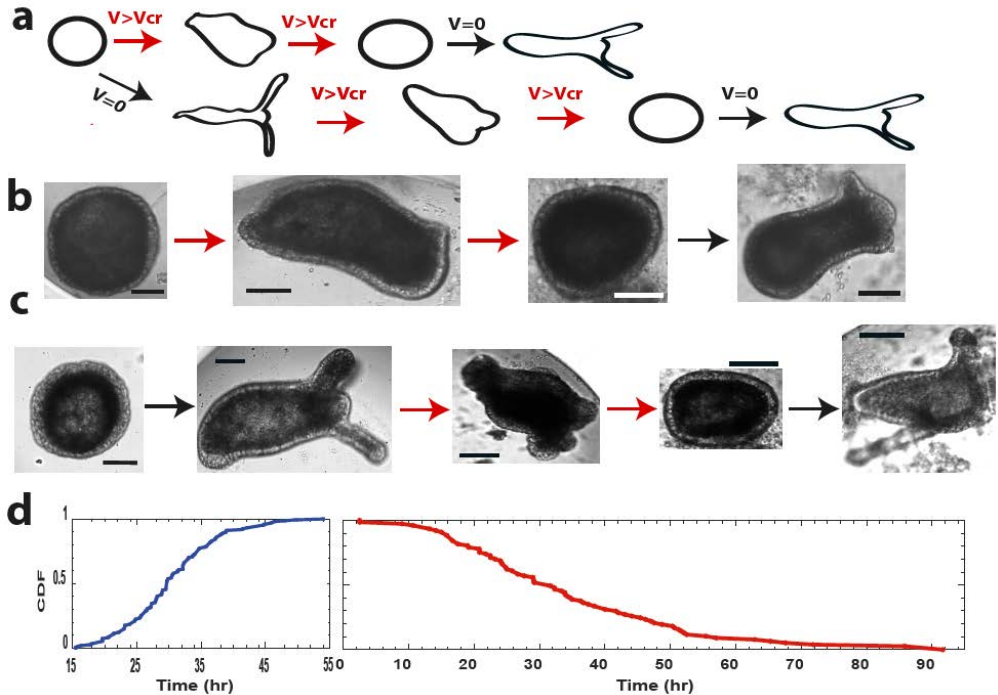

**Fig. 1: Halt and reversal of morphogenesis under external electric fields**. (a) Possible trajectories for an incipient spheroid tissue under a field. (b) Images depicting the trajectory scheme in (a) top row: Time (hrs from tissue excising) and voltages (Volts) for the images (from left): 5.5, 0; 42, 18; 68, 25.5; 97, 0. The voltage is switched off at 71 hrs. Bars 100 μm scale. (c) Images depicting the trajectory scheme in (a) bottom row: Time (hrs from tissue excising) and voltages (Volts) for the images (from left): 2, 0; 53, 0; 80, 25; 108, 31; 136, 0. The voltage is switched off at 108 hrs. Bars 100 μm scale. (d) Cumulative statistics for regeneration and reversal of morphogenesis. (left) Cumulative statistics (147 tissue samples from 15 different experiments) exhibiting their first regeneration in the absence of an external voltage or a voltage below the critical value. Time is measured from the point of tissue excision and regeneration is identified as the emergence of tentacles. All samples regenerated between 15-55 hrs, in agreement with previous results [25]. (right) Cumulative statistics for reversal of morphogenesis, the folding back of fully regenerated *Hydra* into the incipient spheroid morphology, for 77 tissue samples (from 15 different experiments). All samples first regenerated into a mature *Hydra*. Folding time is estimated from the point at which the voltage was first set above 15 V, the minimal voltage observed to affect morphology (e.g., shortening of the tentacles).

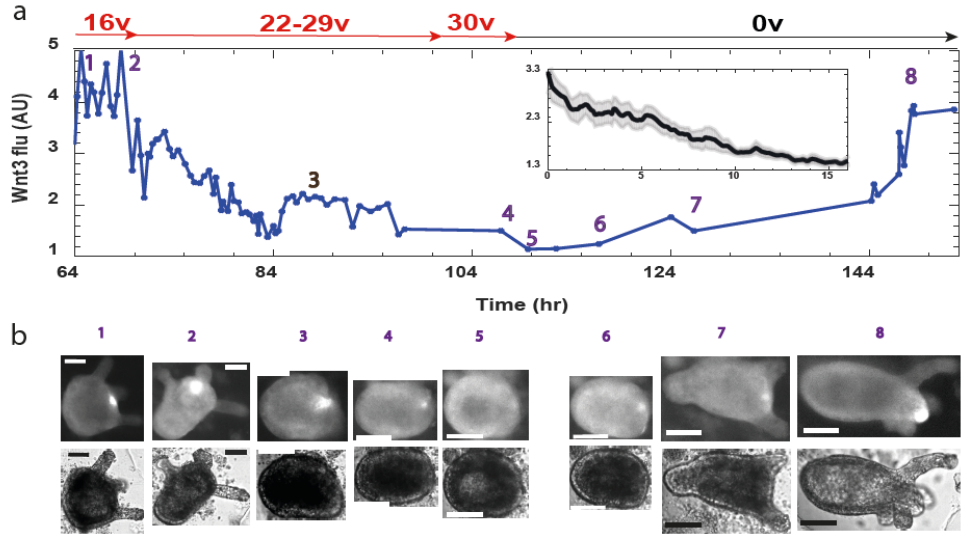

**Fig. 2: The decay of *Wnt3* activity upon reversal of morphogenesis.** Transgenic *Hydra* expressing a GFP probe under the control of the *Wnt3* promoter [31, 32], imaged at 1 min time-lapse fluorescence microscopy. (a) Example trace of the *Wnt3* activity from a tissue sample, at different time points (from the tissue excision time), estimated from the average fluorescence density of the GFP signal in a small region around the center of the signal, relative to the average GFP signal in the background tissue. The curve measures the trajectory of the *Wnt3*-activity decay, as the voltage is increased (Voltage values marked in red), and its recovery upon the renewal of regeneration. The tissue first regenerated into a mature *Hydra* in the absence of an external field, expressing a clear GFP signal at the head organizer in the tip of the head (hypostome). An external field above the critical voltage, leads to reversal of morphogenesis and the decay of the *Wnt3* signal (1-5). Resumption of regeneration leading to a mature *Hydra* and the reemergence of the *Wnt3* signal after the voltage is switched off (6-8). Inset: the mean decay of the GFP marking the *Wnt3* activity, averaged from 7 tissue samples (the one in the main figure plus other 6 shown in Fig. S3; from 5 different experiments). The different curves are first shifted to time zero at the peak of the signal, then interpolated to give the same sampling and averaged. The shadow around the curve marks the standard errors estimated from these measurements. (b) Fluorescence images of the GFP channel (upper row) and the corresponding bright-field images (lower row) at time points marked in (a). Bars 100 μm scale.

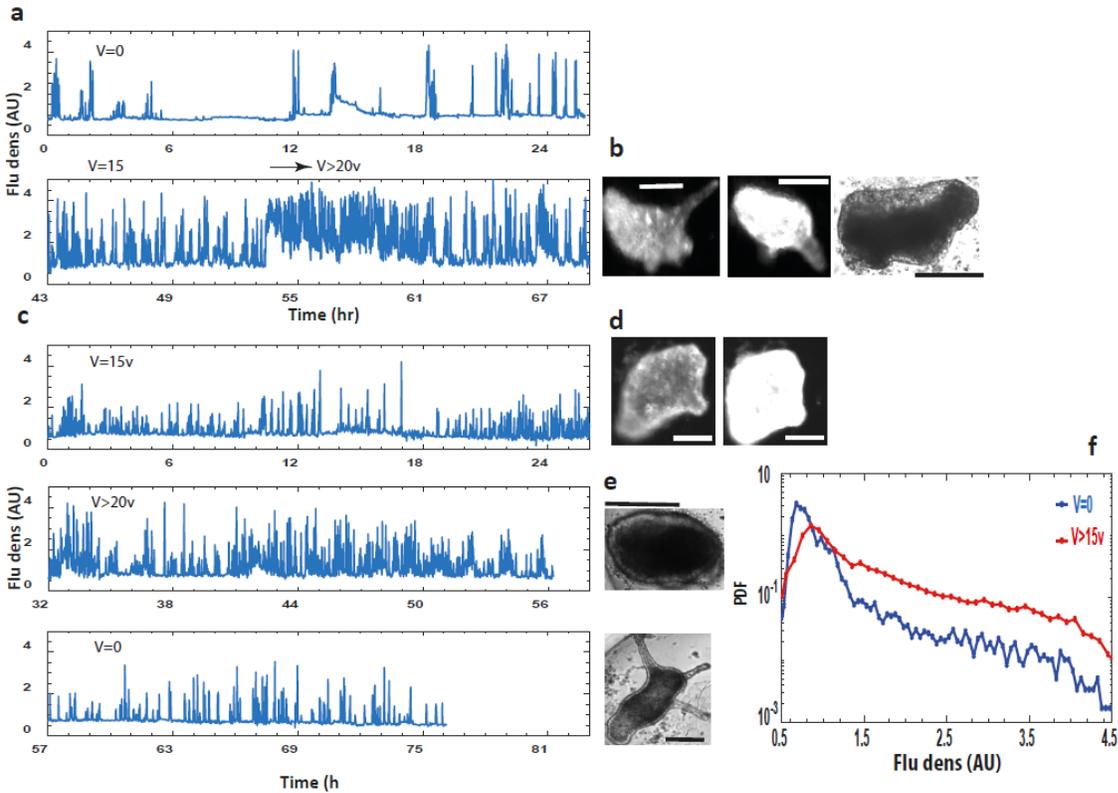

**Fig. 3: Calcium dynamics.** A transgenic *Hydra* expressing the fast $Ca^{2+}$ fluorescence probe [30, 35] in its epithelial cells is utilized. (a) $Ca^{2+}$ dynamics as measured by the fluorescence density (fluorescence signal per unit area) at V=0 (upper trace) and at high external voltage (lower trace). The measurement starts 3 hrs after the tissue excision. The lower trace starts at 15 V, which is below the critical value, and the voltage is then increased to higher values (25 V; arrow). (b) Fluorescence images of the sample (at the end of the trace), at the baseline and high activity (two left images: 69 hrs from excision, 25 V) and a BF image (right; 79 hrs, 25 V). (c) $Ca^{2+}$ activity (as in (a)) of a tissue under 15 V at the onset of the measurement (upper trace). Voltage is increased to 24 V (middle trace) and then switched off (lower trace). (d) Fluorescence images of low and high $Ca^{2+}$ activity at the end of the upper trace (29 hrs, 15 V). Note the tentacles indicating that the tissue regenerated. (e) BF images showing reversal of the morphology at the end of the middle trace, and resumption of regeneration at the end of the lower trace (top image: 63 hrs, 24 V; lower image: 77 hrs, 0 V). (f) Normalized distributions of fluorescence densities at V=0 (blue) and high voltage (>15 V; red). Each curve summarizes statistics of 9 tissue samples (from 3 different experiments, >10,000 measurement points for each curve). Note the y-axis logarithmic scale. All fluorescence measurements were extracted from time-lapse movies at 1 min resolution. The traces show the average fluorescence density of the entire tissue relative to the mean fluorescence density in the background.

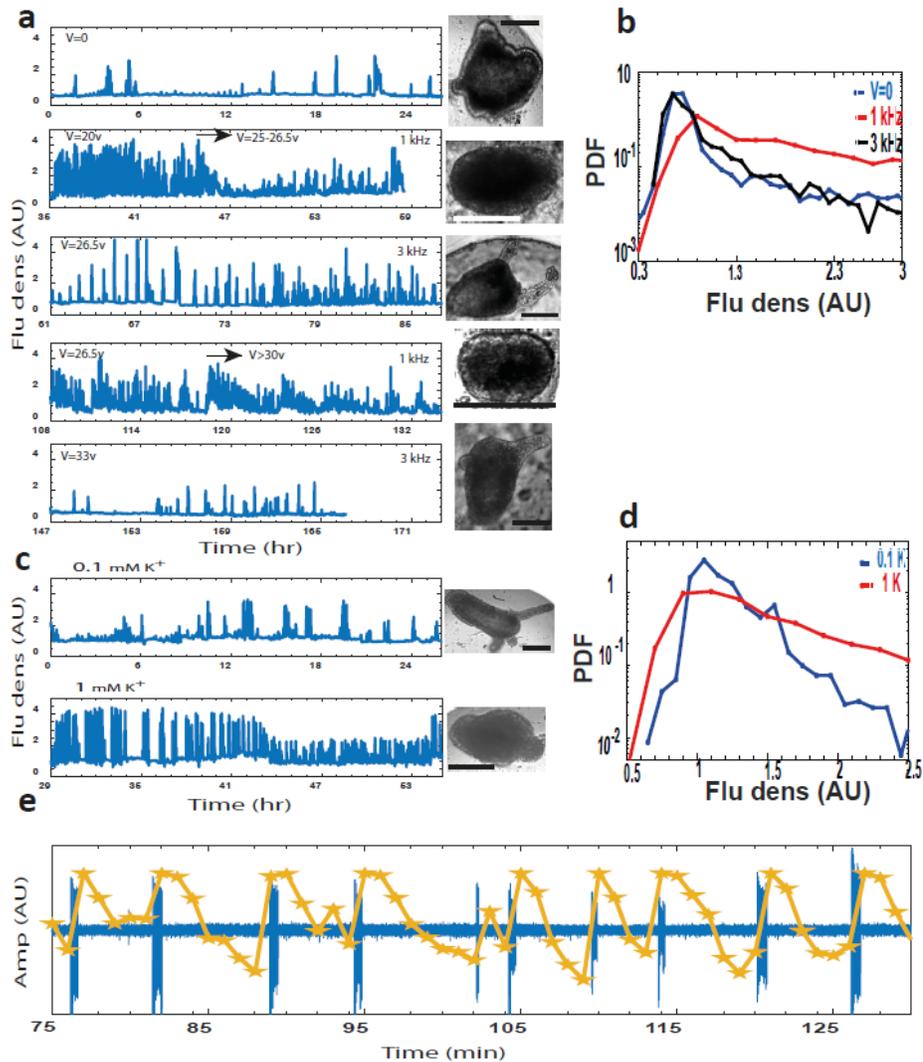

**Fig. 4: Enhanced electrical excitability of the tissue triggering reversal of morphogenesis.** (a) $Ca^{2+}$ fluorescence density for V=0 (first trace) and 20-26.5 V at 1 kHz (second trace). Bf images (at end point of each corresponding trace). (Third trace) 26.5 V, 3 kHz, showing resumption of a fully regenerated *Hydra* (attached image). (b) Normalized fluorescence density distributions comparing: V=0 (blue), high voltage at 1 kHz (red) and 3 kHz (black). Each curve summarizes two tissue samples, with >3000 measured points at 1 min resolution. Note the y-axis logarithmic scale. Repeated frequency switching; 1 kHz at the same voltage (26.5 V; fourth trace). Increased voltage (33 V), leads to reversal of morphogenesis (attached image). A second frequency switch to 3 kHz shows again a reduction in $Ca^{2+}$ activity (fifth trace) and resumption of regenerated *Hydra* (attached image). (c) $Ca^{2+}$ fluorescence density for: 0.1 mM (normal medium; upper trace) and 1 mM (lower trace) $K^+$ medium. Attached BF images at the end point of the corresponding trace. (d) Fluorescence density distributions for 1 mM (red) and 0.1 mM (blue) $K^+$ (2 tissue samples, ~4000 measurement points for each curve). Note the y-axis logarithmic scale. All $Ca^{2+}$ signals shown are measurements of the fluorescence density of the entire tissue relative to the average fluorescence level in the background, estimated from time-lapse imaging at 1 min resolution. (e) Example of a direct measurement of the spontaneous electrical potential by a silver-chloride electrode. Each spontaneously arising $Ca^{2+}$ spike (yellow) is stimulated by a pre-burst of action potentials (blue; Methods). Overall we measured 174 $Ca^{2+}$ spikes in tissue fragments with only 7 among them with no clear burst of electrical spikes (see Figs. S7a-b for more examples and voltage traces).

**Acknowledgements**

We thank Kinneret Keren and Shimon Marom for helpful discussions and comments on the manuscript. We thank Naama Brenner for comments on the manuscript. We thank our lab members: Anton Livshits, Lital Shani-Zerbib, Yonit Maroudas-Sacks and Liora Garion for technical help. Special thanks to Gdalyahu Ben-Yoseph for superb technical help in designing and constructing the experimental setup and in providing the infrastructure enabling the experiments.
We thank: Prof. Thomas Bosch and Dr. Alexander Klimovich for their help in generating the $Ca^{2+}$ strain; Prof. Bert Hobmayer for generously providing the transgenic *Hydra* expressing lifeact; Prof. Brigitte Galliot for generously providing the *Wnt3* strain that was first constructed in Prof. Thomas Holstein lab.
**Funding:** E.B. acknowledged support by the Israel Science Foundation, Grant # 228/17.

## Materials and Methods

### *Hydra strains, culture and sample preparation*

Experiments are carried out with three transgenic strains of Hydra Vulgaris (AEP): A HyWnt3:GFP-HyAct:dsRED transgenic strain (a generous gift from B. Galliot University of Geneva (*[32]*) utilizing the hoTG-HyWnt3FL-EGFP-HyAct:dsRED plasmid from T. Holstein, Heidelberg (*[31]*)); A transgenic line with a GCaMP6s probe reporting $Ca^{2+}$ activity we generated in the Kiel center using a modified version of the pHyVec1 plasmid (Addgene catalog no. 34789) which replaces the GFP sequence with a GCaMP6s sequence that was codon-optimized for Hydra (*[35]*). The embryos were grown and propagated for a few weeks. We selected *Hydra* expressing GCaMP6s in their epithelium cells and propagate them until a stable signal covering the entire animal emerges throughout the population. A third transgenic line expressing lifeact-GFP in the ectoderm (generously provided by Prof. Bert Hobmayer, University of Innsbruck, Austria) (*[25]*), is also used in some experiments, allowing to verify the folding-back of the Hydra into a spheroid by imaging the actin fibers (data not shown). The reversal of morphogenesis phenomenon is found to be similar for all strains. Animals are cultivated in *Hydra* culture medium (1mM NaHCO3, 1mM CaCl2, 0.1mM MgCl2, 0.1mM KCl, 1mM Tris-HCl pH 7.7) at 18°C. The animals are fed every other day with live Artemia nauplii and washed after ~4 hours. Experiments are initiated ~24 hours after feeding.

Tissue segments are excised from the middle of a mature Hydra using a scalpel equipped with a #15 blade. To obtain fragments, a ring is cut into ~4 parts by additional longitudinal cuts. Fragments are incubated in a dish for ~3 hrs to allow their folding into spheroids prior to transferring them into the experimental sample holder. Regeneration is defined as the appearance of tentacles, and the regeneration time is defined as the time interval between excision and the appearance of the first tentacle.

### *Sample holder*

Spheroid tissues are placed within wells of ~1.3 mm radius made in a strip of 2% agarose gel (Sigma) to keep the regenerating *Hydra* in place during time lapse imaging. The tissue spheroid typically of a few hundred μm is free to move within the well. The agarose strip containing 12-13 wells, is fixed on a transparent plexiglass bar of 1 mm height, anchored on a Teflon holder within a 55 mm Petri dish. Two platinum mesh electrodes (Platinum gauze 52 mesh, 0.1 mm dia. Wire; Alfa Aesar) are stretched and fixed by 2% agarose gel from the two sides of the plexiglass bar at a distance of 4 mm, leaving two channels for fluid flow between the electrodes and the samples. The mesh electrodes cover the entire length of the sample holder and their height ensures full coverage of the samples. A peristaltic pump (IPC, Ismatec) flows the medium continuously from an external reservoir (replaced at least once every 24 hrs) at a rate of 170 ml/hr into each of the channels between the electrodes and the samples. The medium covers the entire preparation and the volume in the bath is kept fixed throughout the experiments by pumping medium out from 3 holes which determine the height of the fluid. The continuous medium flow ensures stable environmental conditions and the fixed volume of medium in the bath ensures constant conductivity between the electrodes. All the experiments are done at room temperature.

### *AC generators*

An alternating current (AC) generator (PM5138A, Fluke; or a waveform generator 33621A, Agilent connected to a voltage amplifier A-303, A.A. Lab Systems for voltages above 40 V) is used to set the voltage between the electrodes. The generator is connected directly to one of the electrodes and via a current multimeter (34401A, HP) to the second one, allowing to monitor the current in the system throughout the experiment. The measured current is around 4 mA for applied 10 V and the conductivity is found to be linear with the increase of applied voltage throughout the experiment. The connection to the platinum electrodes is made by sintering 0.4 mm wide platinum wires (Alfa Aesor) to the mesh electrodes, allowing connections to the external devices with only Pt in contact with the medium.

### *Microscopy*

Time lapse bright-field and fluorescence images are taken by a Zeiss Axio-observer microscope (Zeiss) with a 5× air objective (NA=0.25) with a 1.6× optovar and acquired on a CCD camera (Zyla 5.5 sCMOS, Andor). $Ca^{2+}$ fluorescence activity was measured on 696x520 pixels images, by averaging the fluorescence signal over the entire tissue. The $Ca^{2+}$ traces shown in this work depict the fluorescence density averaged over the tissue relative to the background, measured by averaging the fluorescence over a window outside the tissue. The *Wnt3*-GFP fluorescence signal was extracted from the full resolution images of 1392x1040 pixels by measuring the fluorescence density around the center of the GFP signal at the tip relative to the fluorescence density in the background tissue. The sample holder is placed on a movable stage (Marzhauser) and the entire microscopy system is operated by Micromanager, recording images at 1 min resolution.

### *Electrophysiology*

Tissue fragments before folding or tissue spheroids after folding, are immersed in 2% low-gelling agarose (Sigma) covered with a standard medium, placed within a 90 mm petri dish. The measurements are done under a fluorescence microscope (Zeiss Observer), allowing simultaneous fluorescence and electrical measurements. All electrophysiological measurements are done in the absence of external electric fields. Microelectrodes are fabricated by AgCl coated silver wires, threaded in glass capillaries (1.5mm OD and 0.86mm ID, A-M systems). Capillaries are pulled, broken at the extreme tip, and filled with a standard medium solution. A reference electrode made of a thick AgCl coated silver wire is immersed in the dish. The microelectrode is placed on a manipulator and allowed to penetrate the tissue until a stable resting potential is measured. Voltage is measured using AxoPatch 200B amplifier (Axon Instruments) and measurements are inspected by an oscilloscope, as well as digitally acquired at 1000Hz frequency by NI PCI-6259 acquisition card (National Instruments). Time lapse imaging is done at 1 min resolution in bright field and fluorescence channels allowing to image the tissue and record the $Ca^{2+}$ activity.

To extract the recorded spike train, the voltage trace is smoothed (0.1 sec window) and the smoothed trace is subtracted from the original voltage trace. The filtered trace is magnified to emphasize the spikes. The $Ca^{2+}$ activity recorded by the time-lapse fluorescence signal is extracted from each frame within a circle smaller than the tissue fragment to avoid edge effects. The $Ca^{2+}$ signal is manually adjusted to the electrical recording by monitoring the precise starting point of the time-lapse measurement.

**Supplementary Figures:**

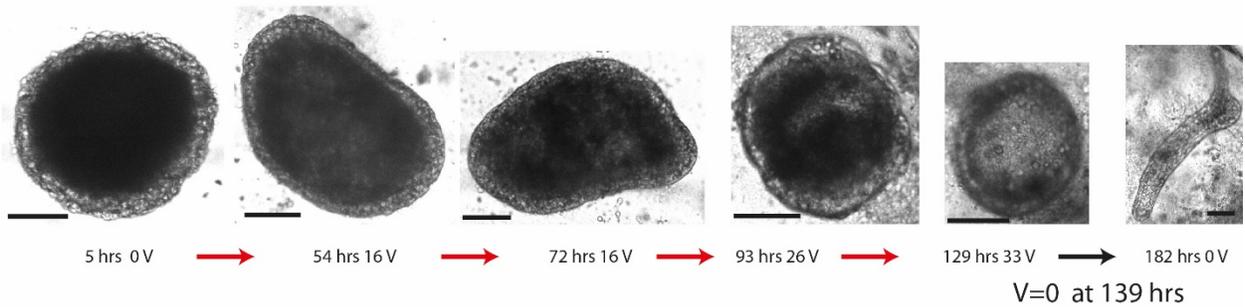

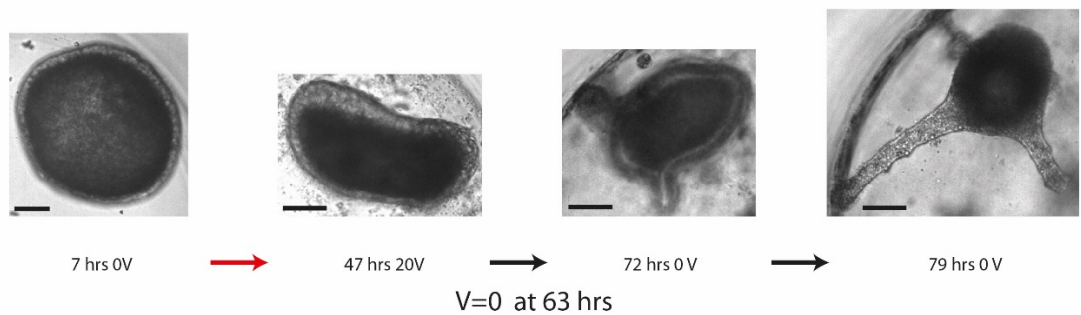

**Fig. S1: Examples of halted regeneration by an external electric field.** A series of images from two experiments (a, b) depicting the trajectory scheme in Fig. 1a top row of the main text. Time (hrs from tissue excising) and voltages (Volts) are indicated for each image. The tissues do not develop under the electric field above its critical value (specific for each tissue; 1 kHz) until time points extending the maximal regeneration time observed in our experiments (~55 hrs) and they readily regenerate into a mature *Hydra* upon switching the voltage to zero at the indicated time. Bars 100 μm scale.

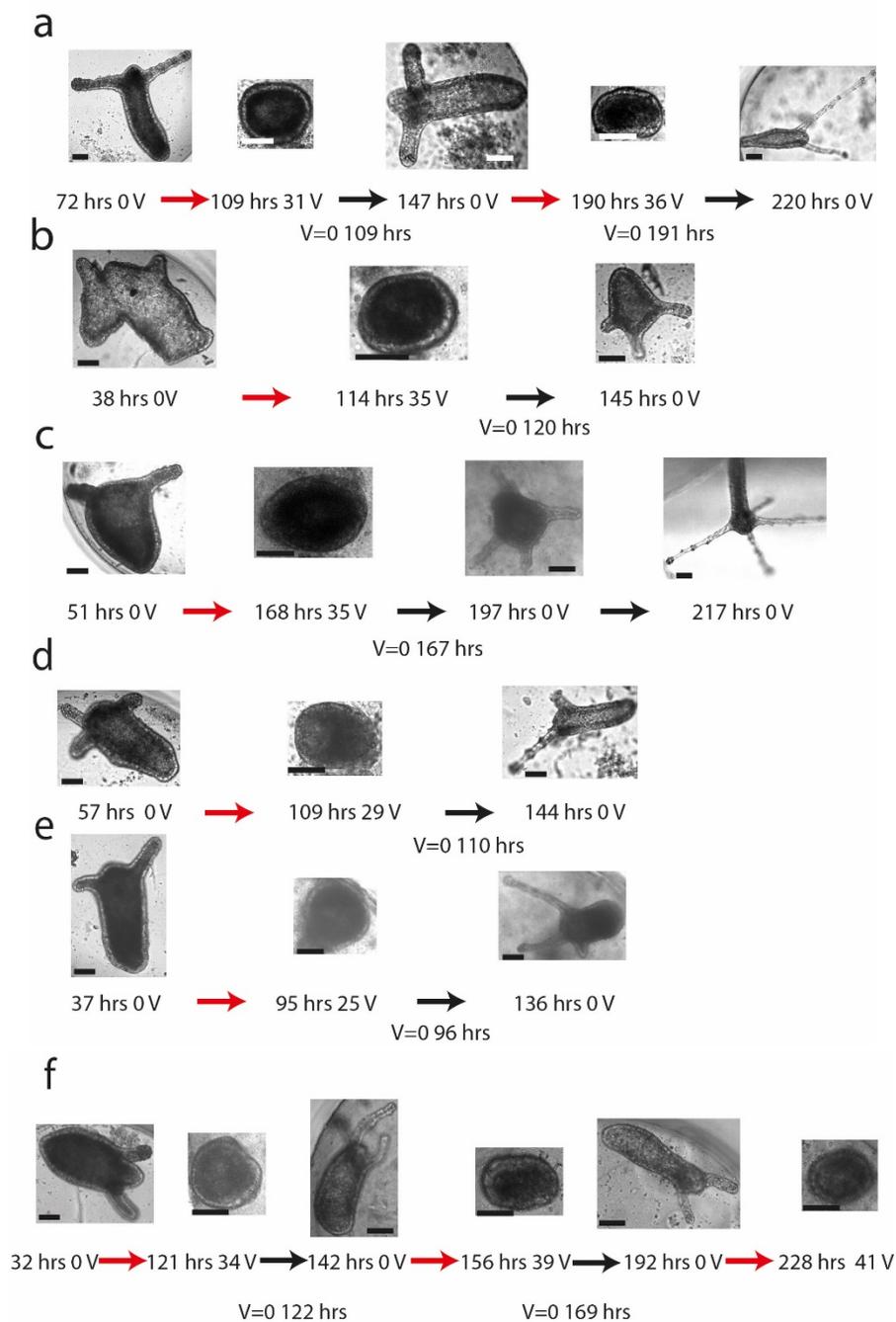

Fig. S2: **Examples of reversal of morphogenesis under an external electric field and the emergence of new morphologies upon renewal of regeneration.** A series of images depicting the trajectory scheme in Fig. 1a bottom row of the main text. Each row of images (a-f) is a separate experiment. Time (hrs from tissue excising) and voltages (Volts) are depicted below the images (external AC field at 1 kHz). In all the experiments presented here, the spheroid tissue first regenerates into a mature *Hydra* and then folds back into a spheroid upon the increase of the externally applied voltage above the critical value (specific for each tissue). The reversed tissue regenerates again upon the reduction of the external voltage to zero at the indicated time. Note that the emerged renewal morphology is in some of the cases not similar to the initial morphology. The images in (a,f) demonstrate that this backward-forward cycle of morphogenesis can be repeated for the same tissue. Bars 100 μm scale.

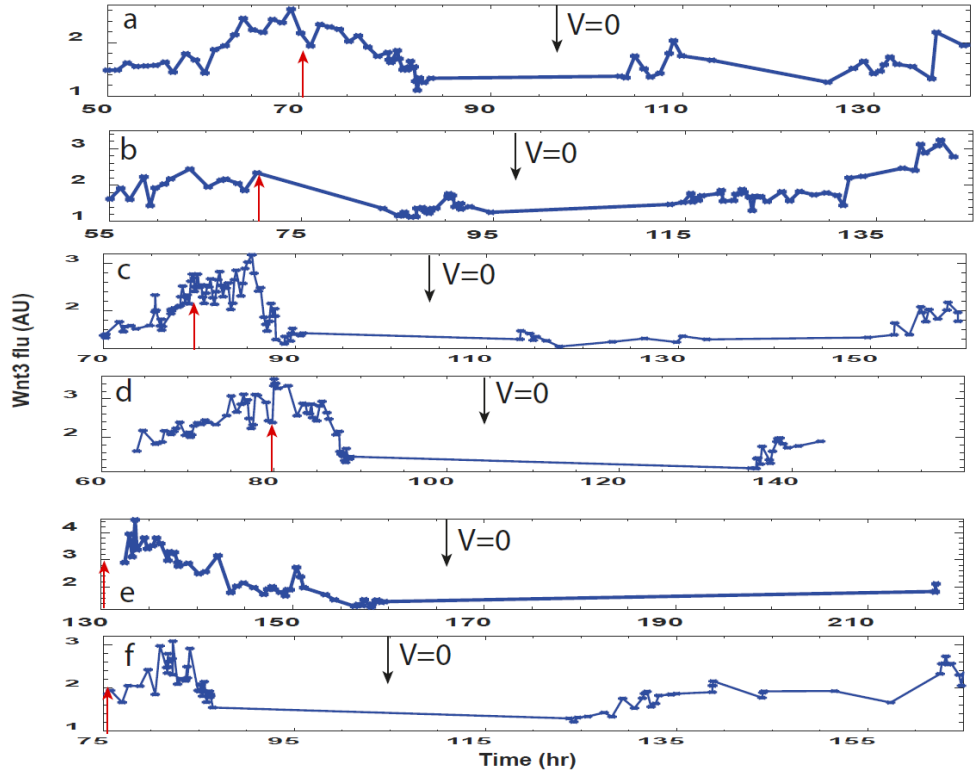

**Fig. S3a: Traces of the *Wnt3* activity under a field.** (a-f) The *Wnt3* activity is estimated in 6 samples (utilized together with the trace shown in Fig. 2 of the main text to produce the mean curve shown in the inset there). Transgenic *Hydra* expressing a GFP probe under the control of the *Wnt3* promoter is imaged under a fluorescence microscope with time-lapse images taken every 1 min (as in Fig. 2 in the main text). The traces show the GFP fluorescence density in the region around the head organizer normalized by the average background GFP level of the tissue outside this region. Measurements at nearby time points show the fluctuation levels of the signal, due to the precise location of the head organizer relative to the surrounding tissue and fluctuations in the spatial organization of the cells comprising it. The red arrows in (a-d) mark the time point at which the external voltage is increased above 20 V (this point is prior to the first measured point for traces (e-f)). The black arrows mark the time points at which the external voltage is switched to zero.

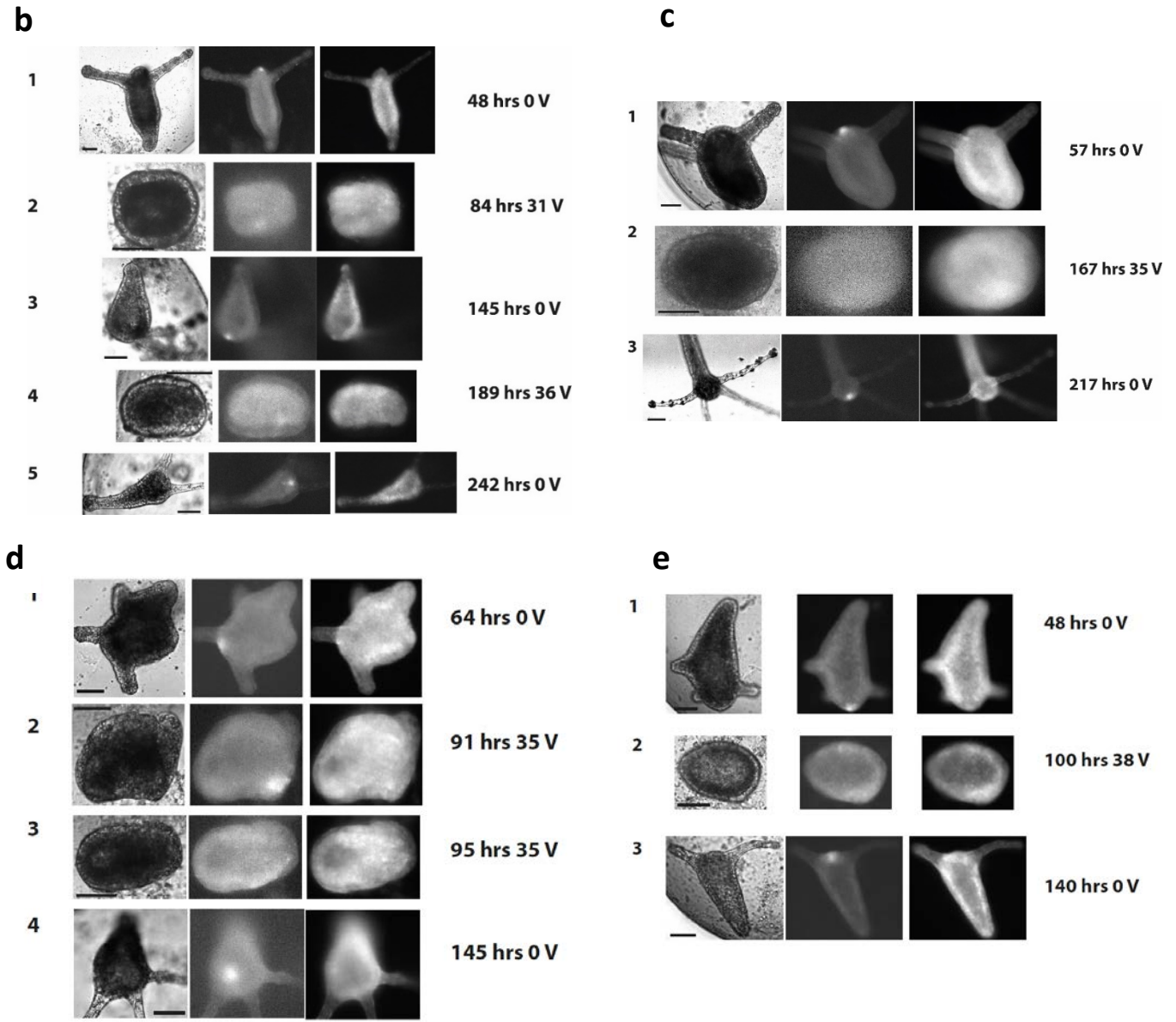

**Figs. S3b-e: Example images of the *Wnt3* activity decay upon reversal of morphogenesis.** The transgenic *Hydra* used to measure the *Wnt3* activity is also expressing dsRED (RFP) under the control of the ubiquitous actin promoter, serving as a fluorescence reference. Four experiments are shown as a sequence of 3 images (from left: bright field, GFP, RFP) over time. The running time (hrs from the point of tissue excision) and the applied external voltage are depicted to the right of the images. A clear fluorescence signal of GFP under the *Wnt3* promoter at the head of a transgenic *Hydra* emerges after the initial spheroid tissue regenerates (1) and decays upon the reversal of morphogenesis in (2;3 in d). The *Wnt3*-activated fluorescence reemerges upon renewal regeneration after the voltage is switched off in (3;4 in d). The experiment in (a) demonstrates a second cycle of reversal morphogenesis (4) and renewal of regeneration leading to re-emergence of the *Wnt3*-activity fluorescence signal. The images correspond to the following traces in (Fig. S3a): (b)-trace f, (c)-trace e, (d)-trace c, (e)-trace a. Bars 100 µm scale.

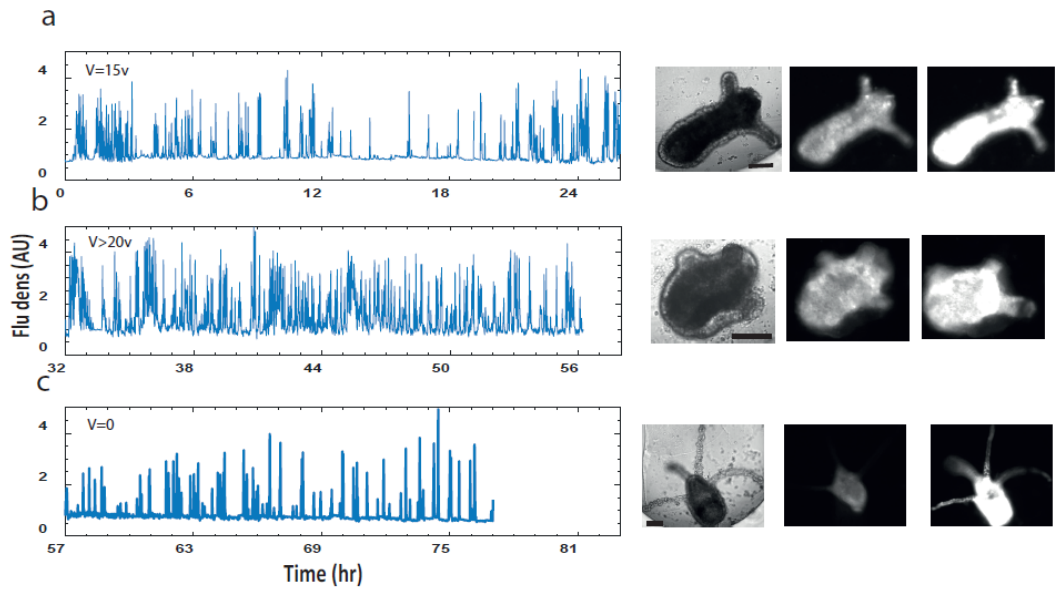

**Fig. S4: Calcium dynamics.** (a) $Ca^{2+}$ dynamics as measured by the fluorescence density (fluorescence signal per unit area of the entire tissue normalized by the background signal) as in Fig. 3c in the main text. The measurement starts 3 hrs after the tissue excision at 15 V which is below the critical value (a). The voltage is then increased to higher values (24 V) in (b) and then switched off in (c). The images on the right show microscopy images of the sample at the end point of each trace (from left: bright field, low and high $Ca^{2+}$ activity depicted by the fluorescence images). Bars 100 μm scale.

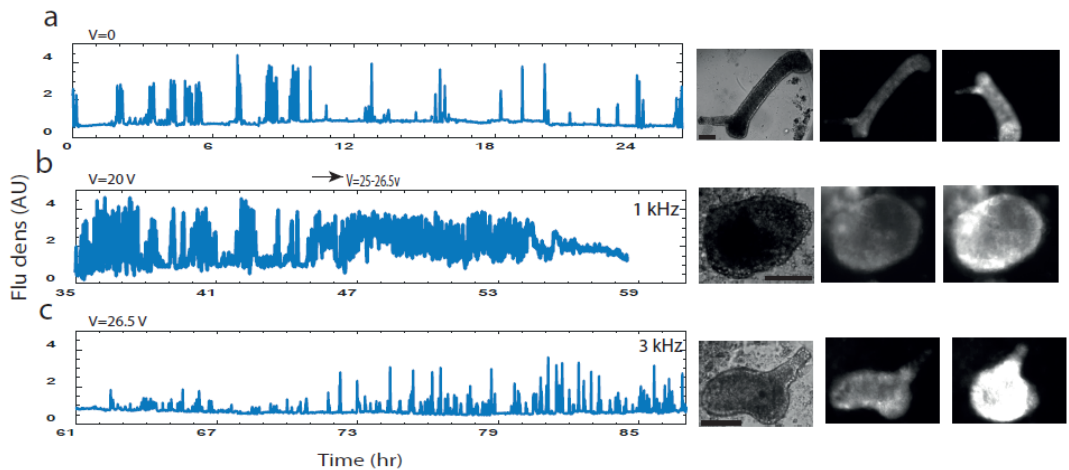

**Fig. S5: Frequency cutoff.** $Ca^{2+}$ activity measured by the fluorescence density (measured over the entire tissue and normalized by the background signal) as in Fig. 4a in the main text, for two different frequencies of the external electric field. The measurement starts 3 hrs after the tissue excision at V=0 (a) and continues at high voltage (20-26.5 V) at 1 kHz (b), showing enhanced $Ca^{2+}$ activity. Switching the frequency of the external AC field to 3 kHz, while maintaining the voltage amplitude at 26.5 V, shows reduction in $Ca^{2+}$ activity (c). The images at the right show microscopy images of the sample at the end point of each trace (from left: bright field, low and high $Ca^{2+}$ activity depicted by the fluorescence images). They show that the tissue fully regenerates for V=0 and then folds back into a spheroid morphology at high voltage, while resumption of a fully regenerated *Hydra* is observed upon the switch of the frequency to 3 kHz. Bars 100 μm scale.

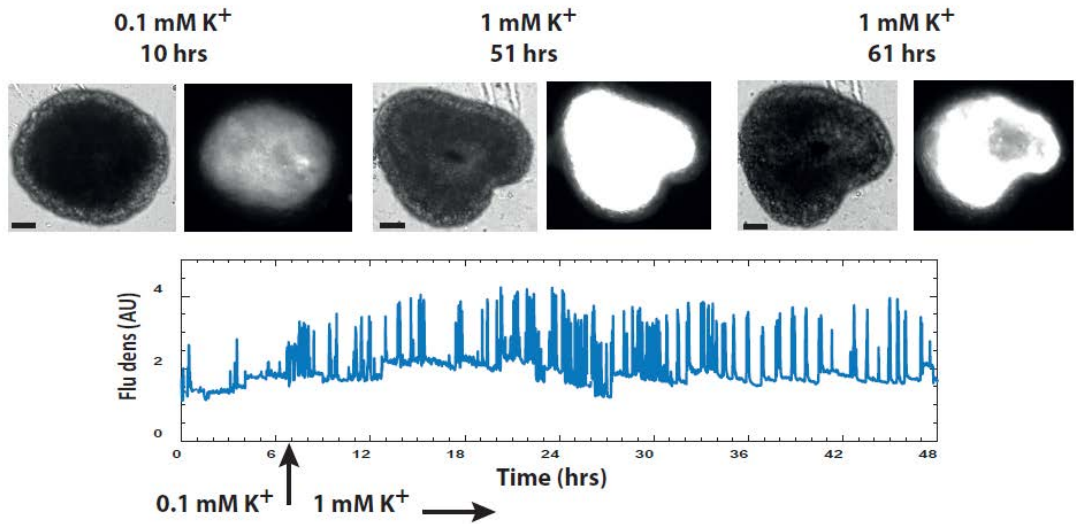

**Fig. S6: Stimulation by elevated potassium.** The trace shows enhanced $Ca^{2+}$ activity measured by fluorescence density (of the entire tissue normalized by the background signal) upon the increase of the potassium ($K^+$) in the medium from 0.1 mM (normal medium) to 1 mM at the time point marked by the arrow. The measurement starts 3 hrs after the tissue excision. Microscopy images of the sample at the indicated time points (hrs from the excision point) and $K^+$ concentrations are shown above the trace in pairs (left: bright field; right $Ca^{2+}$ activity depicted by the fluorescence images). The images show that after 61 hrs there are no signs of regeneration in 1 mM $K^+$ while all samples at the normal *Hydra* medium (0.1 mM $K^+$) regenerate within 55 hrs.

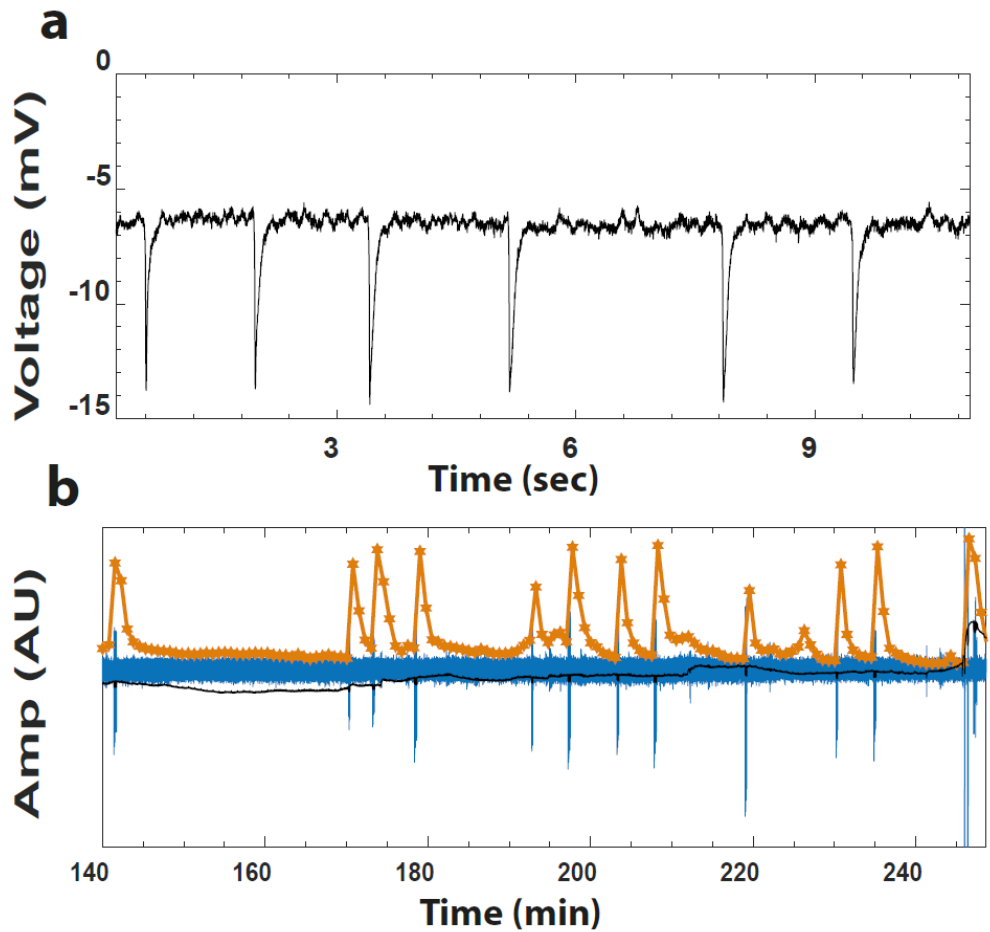

**Fig. S7a: Electrical measurement in a tissue fragment.** (a) Voltage measurement of a spike burst by a silver-chloride electrode inserted into a tissue fragment embedded in a low-melting 2% agarose gel. (b) Simultaneous recording of spontaneous $Ca^{2+}$ activity by fluorescence microscopy (yellow) and the voltage by an electrode (black). The blue trace shows the voltage spikes extracted from the original voltage measurement by subtracting a smoothed trace (0.1 sec window) from the original one and amplifying the resulting trace (distorting the uniphase spikes, but allowing easy identification of them). Each $Ca^{2+}$ spike follows a burst of electrical spikes. The burst in (a) is a zoom over a burst near 190 min in (b). Note the seconds time scale in (a) and minutes in (b). Measurements are done in the absence of an external field.

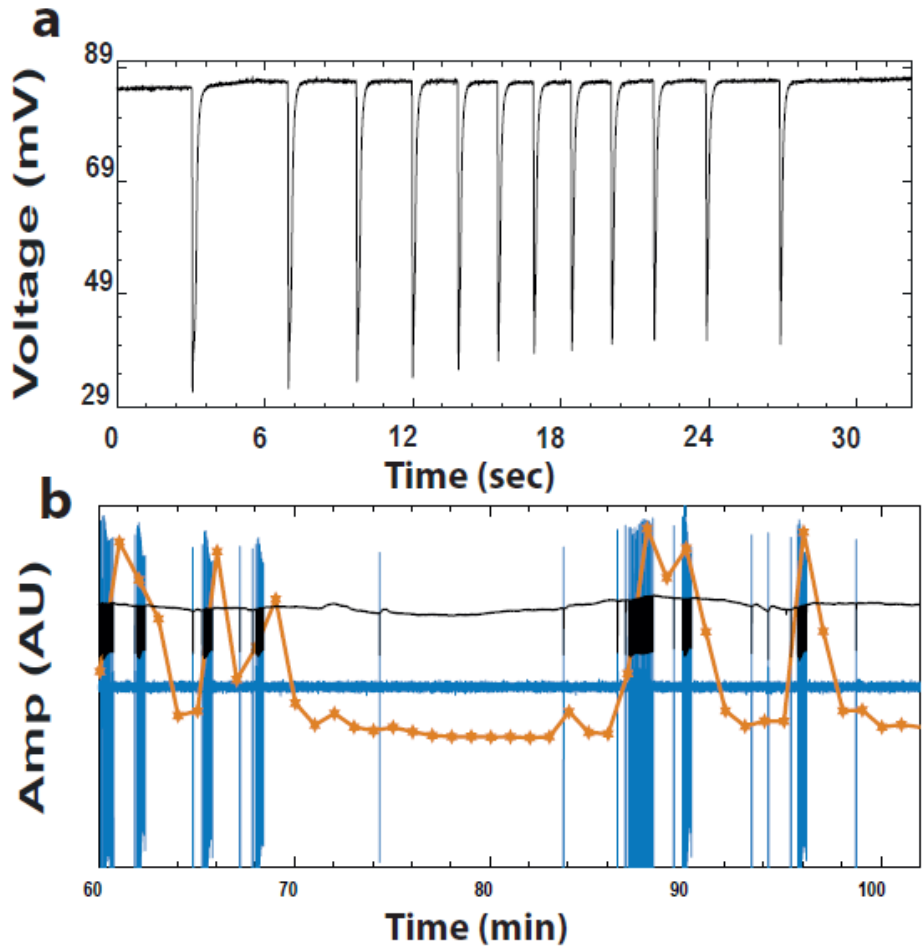

**Fig. S7b: Electrical measurement in a spheroid.** (a) Voltage measurement of a spike burst by a silver-chloride electrode inserted into a tissue spheroid (around 3 hrs after excision and folding in solution) embedded in a low-melting 2% agarose gel. (b) The simultaneously recording of the spontaneous $Ca^{2+}$ activity by fluorescence microscopy (yellow) and the voltage by an electrode (black). The blue trace shows the voltage spikes extracted from the original voltage measurement as in Fig. S7a. Each $Ca^{2+}$ spike follows a burst of electrical spikes. The burst in (a) is a zoom over a burst near 65 min in (b). Note the seconds time scale in (a) and minutes in (b). Measurements are done in the absence of an external field.

**Supplementary Videos:**

**Movie S1: Reversal of morphogenesis.** Bright-field microscopy movie of a tissue spheroid regenerating into a mature *Hydra* and then folding back its morphology under an electric field. The reversed tissue regenerates again upon the reduction of the external voltage to zero. A second round of reversal of morphogenesis then follows. Note the differences in the critical voltages required for reversal of morphogenesis for the first and second rounds; the voltage range in the first round is 20-30 V while in the second round is 30-40 V. The scale of the image is ~1.1 mm and the running time is from the point of tissue excision.

**Movie S2: Decay of *Wnt3* during reversal of morphogenesis.** The same sample as in Movie S1 is observed under a fluorescence microscope for GFP (left) and RFP (right). The GFP is under the *Wnt3* promoter and the RFP is under the control of the ubiquitous *Hydra* actin promoter and serves as a fluorescence reference. The movie shows the two rounds of reversal of morphogenesis as in Movie 1. The scale of the image is ~1.1 mm and the running time is from the point of tissue excision.

**Movie S3: Dynamics of $Ca^{2+}$ in reversal of morphogenesis under an electric field.** Fluorescence microscopy time-lapse of a strain expressing the GCaMP6s probe reporting $Ca^{2+}$ activity in the epithelial cells. The tissue spheroid regenerates under a voltage below the critical value and then folds back as the voltage is increased. The tissue regenerates again upon switching the voltage to zero. The scale of the image is ~1.1 mm and the running time is from the point of tissue excision.

**Movie S4: Frequency cutoff of the external field.** Bright-field microscopy of a tissue under an external AC electric field at two different frequencies. At 1 kHz and voltage above criticality, the *Hydra* exhibits reversal of morphogenesis and folds into a spheroid while it regenerates again at 3 kHz under the same voltage. The movie shows two rounds of the backward-forward morphogenesis cycle under the frequency switch. The scale of the image is ~1.1 mm and the running time is from the point of tissue excision.